\input{aipcheck}

\documentclass[
    ,final            % use final for the camera ready runs
  ]
  {aipproc}

\layoutstyle{6x9}

\newcommand{\be}{\begin{equation}}
\newcommand{\bea}{\begin{eqnarray}}
\newcommand{\ee}{\end{equation}}
\newcommand{\eea}{\end{eqnarray}}
\newcommand{\bpi}{\begin{picture}}
\newcommand{\bce}{\begin{center}}
\newcommand{\epi}{\end{picture}}
\newcommand{\ece}{\end{center}}

\def\gb{{\Gamma}}

\def\g{\widetilde{\Gamma}}

\begin{document}

\title{New insights on non-perturbative Yang-Mills}


\author{Arlene~C. Aguilar}{
  address={Federal University of ABC, CCNH, \\
Rua Santa Ad\'{e}lia 166, CEP 09210-170, Santo Andr\'{e}, Brazil.}
}

\begin{abstract}

In this talk we review some recent results on the
infrared properties of the gluon and  ghost propagators in pure Yang-Mills theories. These
results are obtained from the corresponding Schwinger-Dyson equation formulated in  a 
special truncation scheme, which preserves gauge invariance. The  presence of massless poles
in the three gluon vertex  triggers the generation of a dynamical gluon mass (Schwinger mechanism in $d=4$), 
which gives rise to an infrared finite gluon propagator and ghost dressing function. 
As a byproduct of this analysis  we calculate the Kugo-Ojima function, required for the 
definition of the non-perturbative QCD effective charge within the pinch technique framework. 
We show that the numerical solutions of these non-perturbative 
equations are in very good agreement with the results of $SU(3)$ lattice simulations. 
  
\end{abstract}

\maketitle

%%%%%%%%%%%%%%%%%%%%%%%%%%%%%%%%%%%%%%%%%%%%
%% MAINMATTER
%%%%%%%%%%%%%%%%%%%%%%%%%%%%%%%%%%%%%%%%%%%%

In the last three decades we have accumulated ample experimental evidence
corroborating our conviction  
that QCD is indeed the theory of the strong interactions. Most of these experimental
data come from short distance processes, whose theoretical  description can be performed  
by means of perturbation theory. The reason
why perturbation theory describes so successfully the experimental results in the ultraviolet (UV) region is due to    
one of the most intrinsic characteristics of QCD, namely asymptotic freedom \cite{Marciano:1977su}. 
On the other hand, at large distances, another extraordinary phenomenon of the QCD manifests itself: the color confinement,
which prevents the fundamental excitations of the theory (quarks and gluons) from appearing as free particles;
only color singlets are observed as asymptotic states. 

One of the major challenges of the strong interactions is to explain confinement from  first principles. 
One possible way to gain some insights on the nature of the confinement, and  how the infrared (IR) dynamics of the theory  
works, is to study the  Green's functions of the fundamental degrees of freedom, gluons, quarks and ghosts. Even
though it is well-known that these quantities are not physical, since they depend on the 
gauge-fixing scheme and the parameters used to renormalize them, they do capture crucial aspects of the underlying
 perturbative and non-perturbative dynamics. In addition, 
when appropriately combined, they give rise to physical observables.

The most widely employed tools to explore the IR dynamics of the Yang-Mills theories are: 
(i) the \emph{lattice simulation}, where Monte Carlo techniques are used, 
after discretizing space-time and imposing periodic boundary conditions 
on the finite volume, and 
(ii) the \emph{Schwinger-Dyson equations} (SDE), which form an infinite set of integral equations governing the dynamics of the
off-shell QCD Green's functions.

In this talk we will review some of the recent advances on the 
IR behavior of the gluon and ghost propagator in a pure Yang-Mills theory, obtained  
within the truncation scheme of the SDE based on the pinch technique
(PT) \cite{Cornwall:1989gv} and its connection with the background field method (BFM) \cite{Abbott:1980hw}.  We compare our SDE results
with the available lattice data \cite{Cucchieri:2007md,Bogolubsky:2007ud,Sternbeck:2006rd}, and we discuss how these results are best interpreted by  
assuming the generation of an effective gluon mass  \cite{Cornwall:1982zr, Aguilar:2006gr,Aguilar:2008xm}.            
%
%%%%%%%%%%%%%%%%%%%%%%%%%%%%%%%%
%    Figure 1
%%%%%%%%%%%%%%%%%%%%%%%%%%%%%%%
\begin{figure}[!t]
\begin{minipage}[b]{0.45\linewidth}
\centering
%\hspace{-1cm}
\includegraphics[scale=0.33]{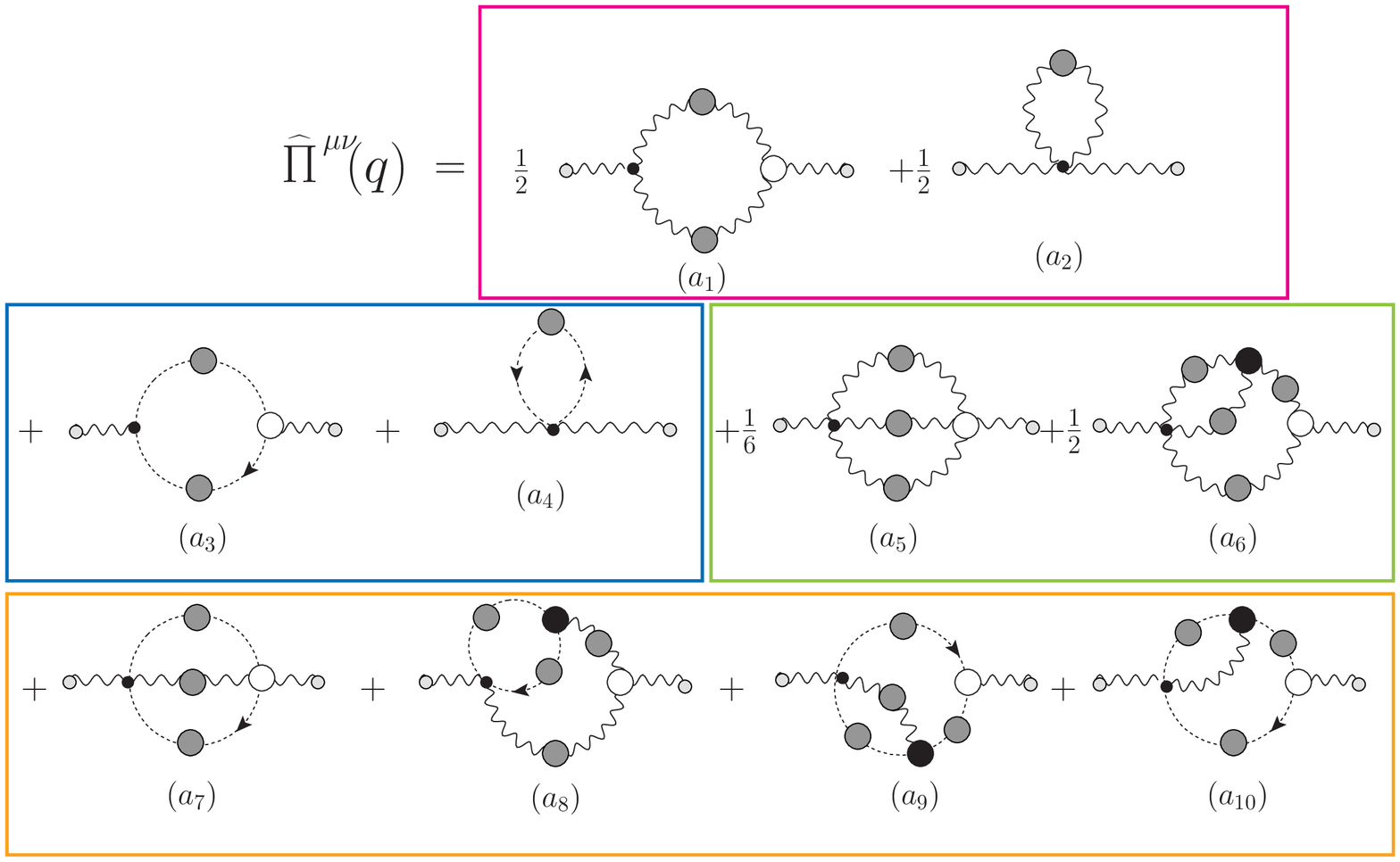}
\end{minipage}
\hspace{0.5cm}
\vspace{1.0cm}
\begin{minipage}[b]{0.50\linewidth}
\centering
\includegraphics[scale=0.5]{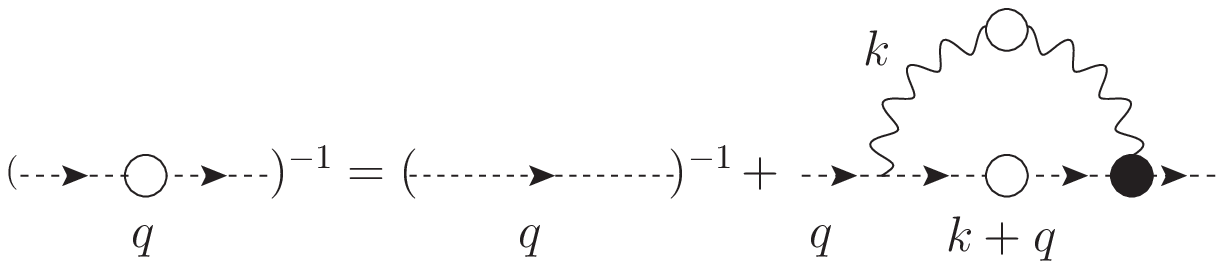}
\includegraphics[scale=0.45]{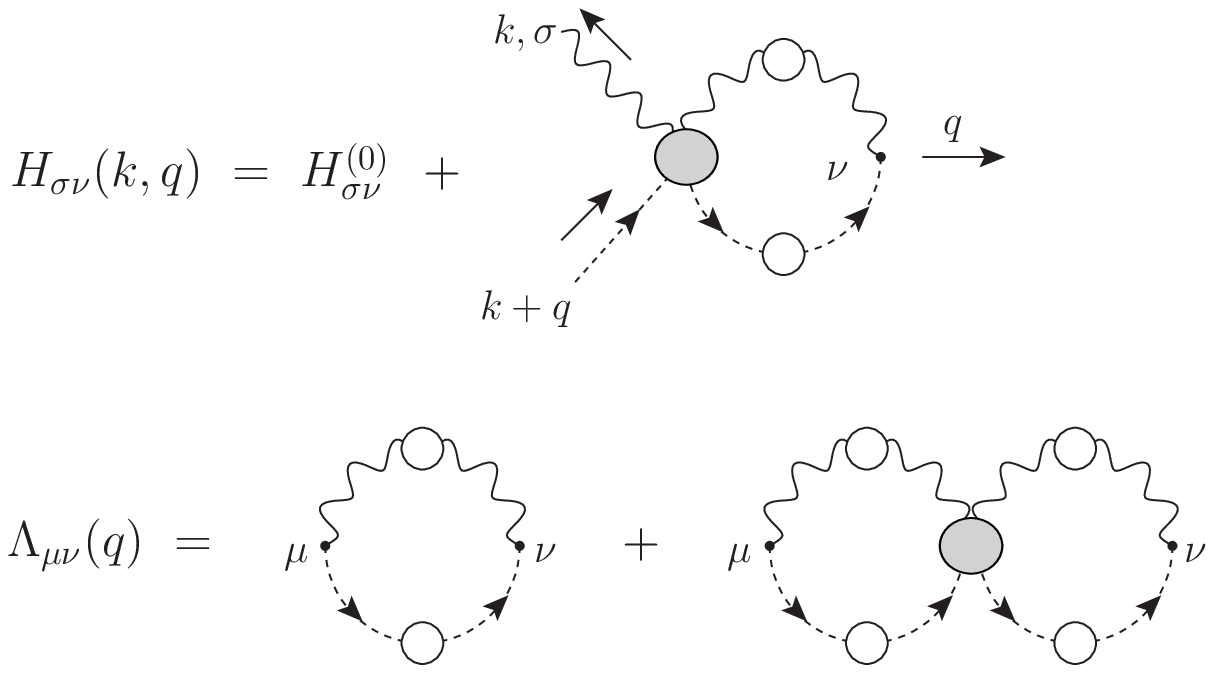}
\end{minipage}
\vspace{-1.0cm}
\caption{The  new SDE for the gluon and ghost propagators, and the auxiliary function in the PT-BFM scheme.}
\label{fig1}
\end{figure}

In the Landau gauge the full gluon propagator, $\Delta_{\mu\nu}(q)$, is transverse, and its 
general form is given by
\be \Delta_{\mu\nu}(q)=-iP_{\mu\nu}(q)\Delta(q^2), \quad \mbox{with} \quad 
P_{\mu\nu}(q)= g_{\mu\nu} - \frac{q_\mu q_\nu}{ q^2}\,,
\label{prop_cov}
\ee
where the scalar function,  $\Delta(q^2)$, is related 
to the all order self-energy  $\Pi_{\mu\nu}(q)=P_{\mu\nu}(q)\Pi(q^2)$ through
\mbox{$\Delta^{-1}(q^2) = q^2 + i \Pi(q^2)$}.
The  full self-energy \mbox{$\Pi_{\mu\nu}(q)$} is expressed in terms of an infinite tower of coupled integral equation 
governing the dynamics of all n-point functions of the theory.
For  practical purposes, this tower of equations, known as gluon SDE, 
can be only employed after devising a self-consistent truncation 
scheme which  selects a  tractable
subset of these equations,  without compromising the crucial
characteristics of the theory. 

Devising  such  a  scheme,  however,  is  very
challenging, especially in the  context of non-Abelian gauge theories,
such as QCD.  One of the main difficult 
resides in the fact that one cannot truncate the conventional gluon SDE,  in any obvious way, without violating 
the basic transversality relation \mbox{$q^{\mu}\Pi_{\mu\nu}(q)=0$}, imposed by the BRST symmetry.  
  
Recently a new truncation scheme, that respects gauge  invariance  at every level  of  the ``dressed-loop''
expansion, has been proposed for the gluon and ghost SDEs \cite{Aguilar:2006gr}. This particular truncation scheme is based on the pinch technique (PT)
\cite{Cornwall:1989gv} and
its correspondence with the background field method (BFM)\cite{Abbott:1980hw} , and 
implements a drastic modification at
the level of the building blocks of the SD series. The PT
enables the construction of new, effective Green's functions
endowed with very special properties; most importantly,
they are independent of the gauge-fixing parameter,
and satisfy QED-like Ward identities (WI) instead of
the usual Slavnov-Taylor identities \cite{Cornwall:1989gv}. Within this formalism  one 
trades the conventional SD series
for another one, written in terms of the new Green's functions,
and then truncates this new series, by keeping only a few
terms in a ``dressed-loop'' expansion, maintaining exact
gauge-invariance. In Fig.~{\ref{fig1}}, we show
the new diagrams composing the  
the PT-BFM self-energy,  to be denoted as  $\widehat{\Pi}_{\mu\nu}(q)$.
Notice that the vertices appearing in the diagrams of Fig.~{\ref{fig1}} are not
the conventional ones, but rather the vertices constructed 
using the Feynman rules of the BFM~\cite{Abbott:1980hw}. 

One of the most interesting properties of $\widehat{\Pi}_{\mu\nu}(q)$
is the way its transversality is enforced. 
More specifically, the SDE is composed of ``one-loop'' and ``two-loop'' dressed blocks that are 
{\it individually transverse}. In fact, the resulting pattern is even more restrictive: 
the gluon and ghost diagrams form separate transverse blocks, represented by  the boxes
in Fig.~{\ref{fig1}}. In this way, we have that the gluonic sector of the  ``one-loop'' dressed blocks satisfies  $q^{\mu}[(a_1)+(a_2)]_{\mu\nu} =0$, while
for the ghost $q^{\mu}[(a_3)+(a_4)]_{\mu\nu} =0$; similarly for the ``two-loop'' blocks, we have  $q^{\mu}[(a_5)+(a_6)]_{\mu\nu} =0$ for the gluonic part, and $q^{\mu}[(a_7)+(a_8)+(a_9)+(a_{10})]_{\mu\nu} =0$ for the ghost sector.
The fact that the transversality is enforced ``block-wise'' allows for a self-consistent truncation of the full gluon SDE \cite{Aguilar:2006gr}.

In addition, the connection between the conventional $\Delta(q^2)$ and the PT-BFM  $\widehat{\Delta}(q^2)$
are  done via a powerful formal identity~\cite{Grassi:1999tp}
stating that 
\be
\Delta(q^2) = 
\left[1+G(q^2)\right]^2 \widehat{\Delta}(q^2), 
\label{bqi2}
\ee
with $G(q^2)$ defined from $\Lambda_{\mu \nu}(q)$, Fig.\ref{fig1}, 
\bea
\Lambda_{\mu \nu}(q) =g_{\mu\nu} G(q^2) + \frac{q_{\mu}q_{\nu}}{q^2} L(q^2) =-iC_{\rm {A}}g^2
\int_k H^{(0)}_{\mu\rho}
D(k+q)\Delta^{\rho\sigma}(k)\, H_{\sigma\nu}(k,q)\,,
\label{LDec}
\eea
where $C_{\rm {A}}$ is the Casimir eigenvalue of the adjoint representation
[$C_{\rm {A}}=N$ for $SU(N)$], and \mbox{$\int_{k}\equiv\mu^{2\varepsilon}(2\pi)^{-d}\int\!d^d k$}, 
with $d=4-\epsilon$ the dimension of space-time.
The vertex $H_{\mu\nu}$, appearing in Eq.~(\ref{LDec}), is related to the full gluon-ghost vertex, $\Gamma_{\nu}(k,q)$, by the
STI $q^\mu H_{\mu\nu}(k,q)=-i\Gamma_{\nu}(k,q)$, and its tree-level counterpart is given by $H_{\mu\nu}^{(0)} = ig_{\mu\nu}$.

It is important to keep in mind that the auxiliary function $G(q^2)$, appearing in 
the definition of $\Lambda_{\mu \nu}(q)$,   plays an instrumental role in the PT-BFM framework, 
since only with it we are able to connect the $\Delta(q^2)$ and  $\widehat{\Delta}(q^2)$.
Interestingly enough, and in the Landau gauge only, $G(q^2)$ coincides with the so-called Kugo-
Ojima (KO) function; this latter function, and in particular its value in the deep IR, is
intimately connected with the corresponding well-known confinement criterion  \cite{Kugo:1979gm, Aguilar:2009pp}.

In one loop dressed approximation, the PT-BFM self-energy is given by \mbox{$\widehat{\Pi}_{\mu\nu}(q) = [(a_1)+(a_2)+(a_3)+(a_4)]P_{\mu\nu}$(q)}, and
using the identity (\ref{bqi2}), we can express the gluon SDE of  Fig.~\ref{fig1} 
as an integral equation involving only $\Delta(q^2)$,  in the following way \cite{Aguilar:2008xm} 
\be
\Delta^{-1}(q)= 
\frac{q^2  + i[(a_1)+(a_2)+(a_3)+(a_4)]}{[1+G(q)]^2} \,. 
\label{SDgl}
\ee

Moreover, as shown in Fig.~\ref{fig1}, the ghost SDE is the same as in the conventional formulation, namely 
\be
iD^{-1}(q) = q^2 + i C_A g^2 \int_k
\Gamma^{\mu}\Delta_{\mu\nu}(k)\gb^{\nu}(k,q) D(q+k),
\label{SDgh}
\ee
where $\Gamma_{\mu}$ is the standard (asymmetric) gluon-ghost vertex at tree-level,
and $\gb^{\mu}(k,q)$ its fully-dressed counterpart, with $k$ representing the 
momentum of the gluon and $q$ the one of the outgoing ghost.

Next, we use for the two vertices appearing in Eq.(\ref{LDec}) and (\ref{SDgh}) their tree-level values,
\mbox{$H_{\mu\nu}(k,q)=ig_{\mu\nu}$}, and \mbox{$\gb_{\mu}(k,q) = -q_{\mu}$} respectively. Then, setting 
\mbox{$f(k,q) \equiv  (k \cdot q)^2/{k^2 q^2}$}, one may show that \cite{Aguilar:2009nf} 
\bea
F^{-1}(q^2) &=& Z_c +g^2 C_{\rm {A}} \int_k [1-f(k,q)] \Delta (k)  D(k+q),
\nonumber\\
1+G(q^2) &=& Z_c + \frac{g^2 C_{\rm {A}}}{d-1}\int_k [
(d-2)+ f(k,q)]\Delta (k)  D(k+q),
\nonumber\\
L(q^2) &=& \frac{g^2 C_{\rm {A}}}{d-1}\int_k 
[1 - d \,f(k,q)]\Delta (k)  D(k+q)\,,
\label{simple}
\eea 
where $F(q^2)$ is the dressing function of the ghost propagator defined as \mbox{$D(q^2)= iF(q^2)/q^2$}.

It is important to mention that there exists a powerful formal identity relating $F(q^2)$, $G(q^2)$, and $L(q^2)$, namely  
$F^{-1}(q^2) = 1+G(q^2)+L(q^2)$ ~\cite{Grassi:2004yq}.

In addition to its formal derivation~\cite{Grassi:2004yq}, 
the above relation has been recently obtained at the level of the SDEs defining these 
three quantities~\cite{Aguilar:2009nf}. Adding the three equations of (\ref{simple}) we can verify 
that above identity is indeed satisfied under the approximations employed.

One of the most crucial ingredients of this truncation scheme is
the Ansatz employed for the PT-BFM vertex ${\g}_{\mu\alpha\beta}(q,k_1,k_2)$, 
appearing in the diagram $(a_1)$ of Fig.\ref{fig1},  
\be
{\g}_{\mu\alpha\beta}= \Gamma_{\mu\alpha\beta}^{(0)} + i\frac{q_{\mu}}{q^2}
\left[\Pi_{\alpha\beta}(k_2)-\Pi_{\alpha\beta}(k_1)\right]\,,
\label{gluonv}
\ee
whose essential feature 
is the presence of massless 
pole terms, required  for triggering the Schwinger mechanism.
A very detailed discussion about the properties of the 
above Ansatz can be found in ~\cite{Aguilar:2008xm}. 
After using the vertex of Eq.(\ref{gluonv}) in the gluon SDE of (\ref{SDgl}),
we obtain a lengthy equation  that we do not report here. 

\begin{figure}[!t]
\begin{minipage}[b]{0.45\linewidth}
\centering
%\hspace{-1cm}
\includegraphics[scale=0.29]{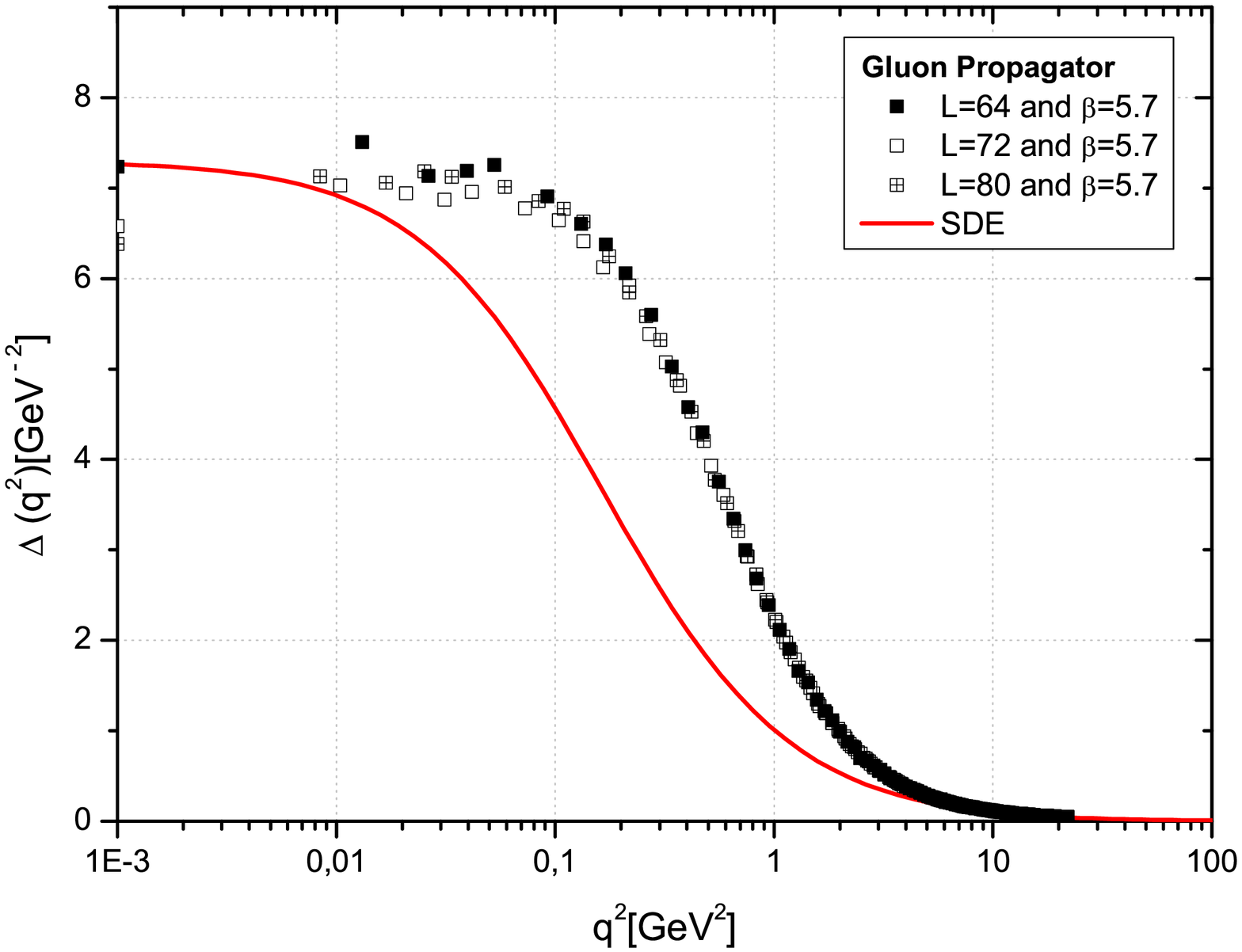} \\
\includegraphics[scale=0.30]{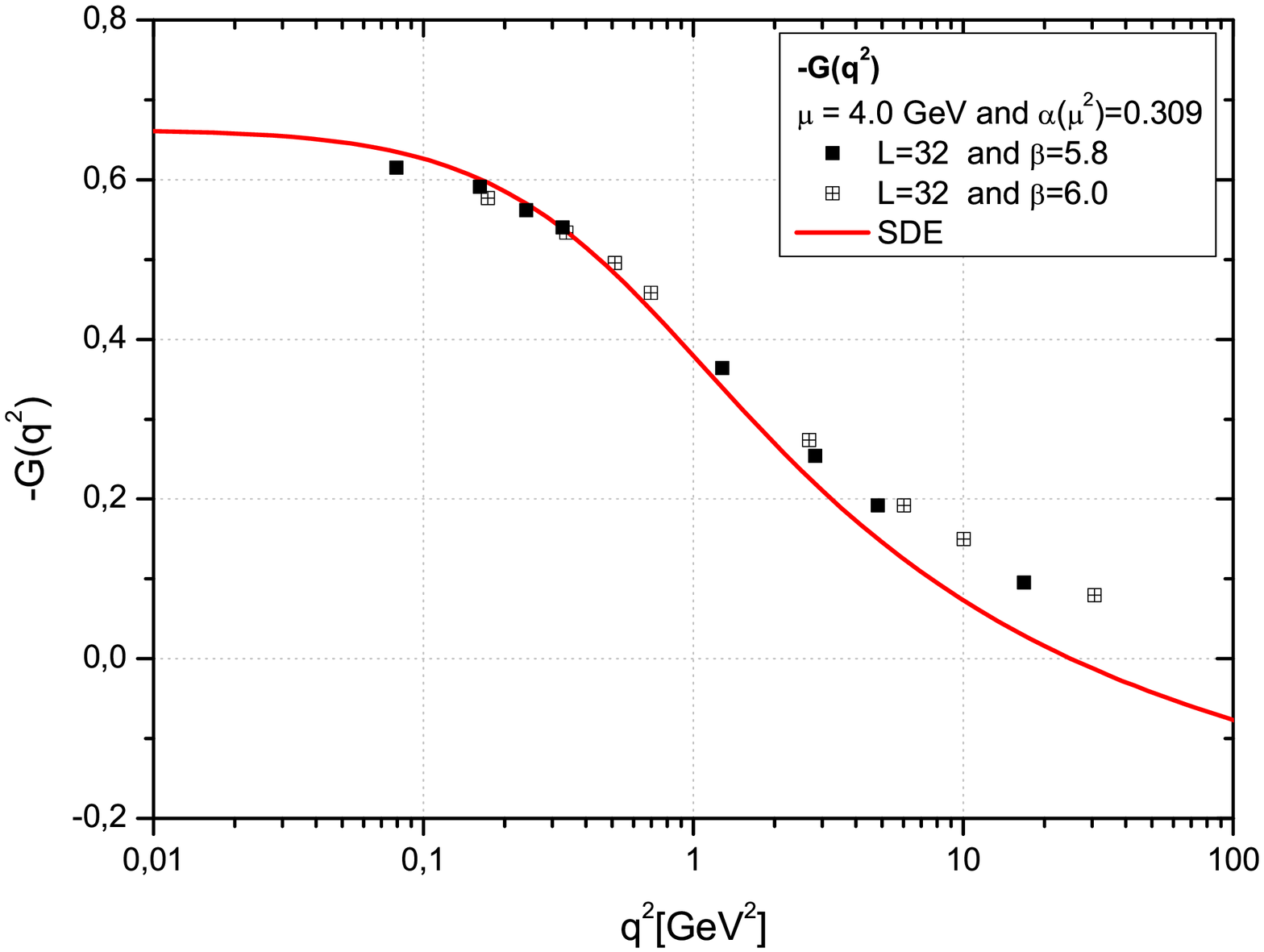}
\end{minipage}
\hspace{0.5cm}
\begin{minipage}[b]{0.50\linewidth}
\centering
%\hspace{-1.5cm}
\includegraphics[scale=0.30]{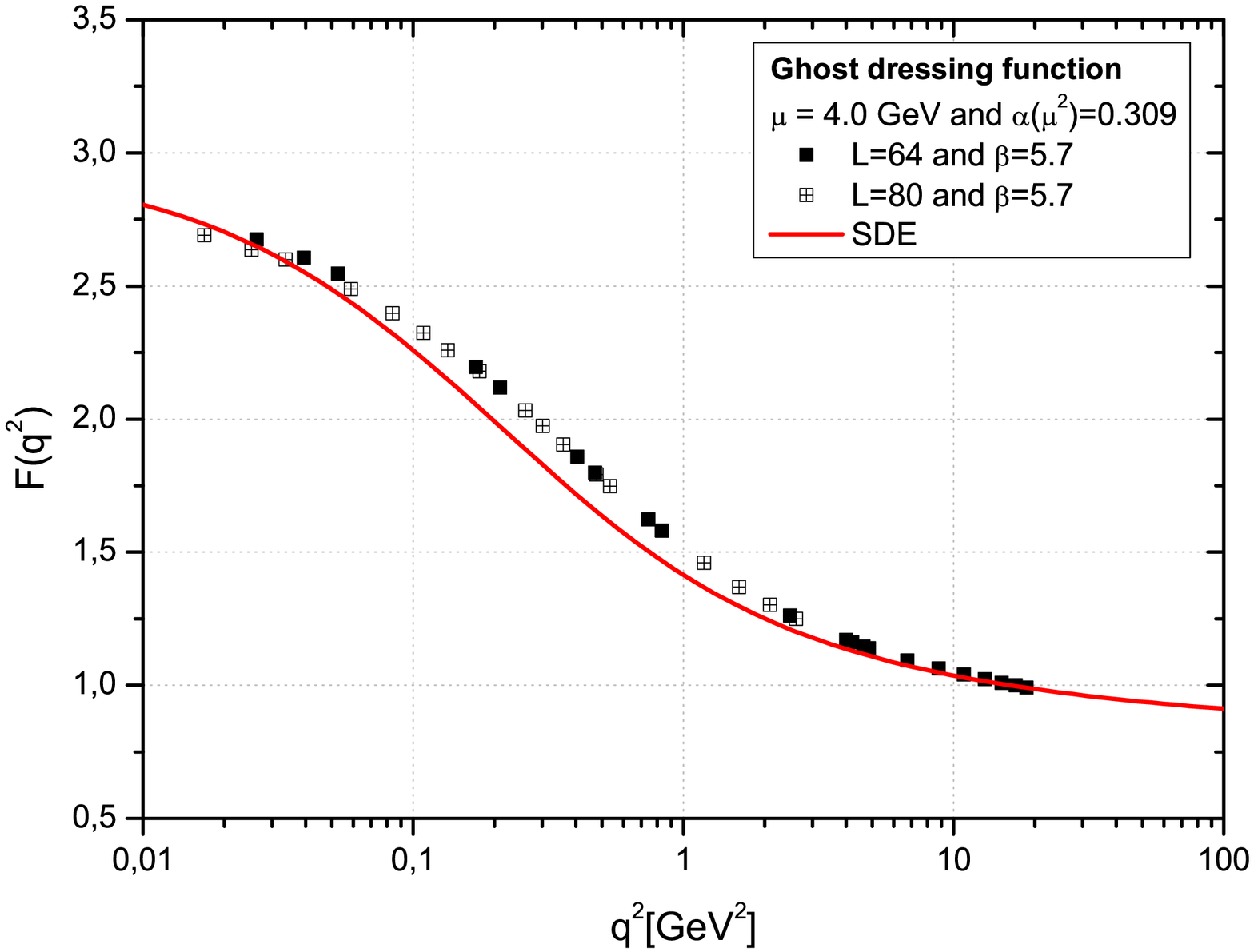} \\
\includegraphics[scale=0.30]{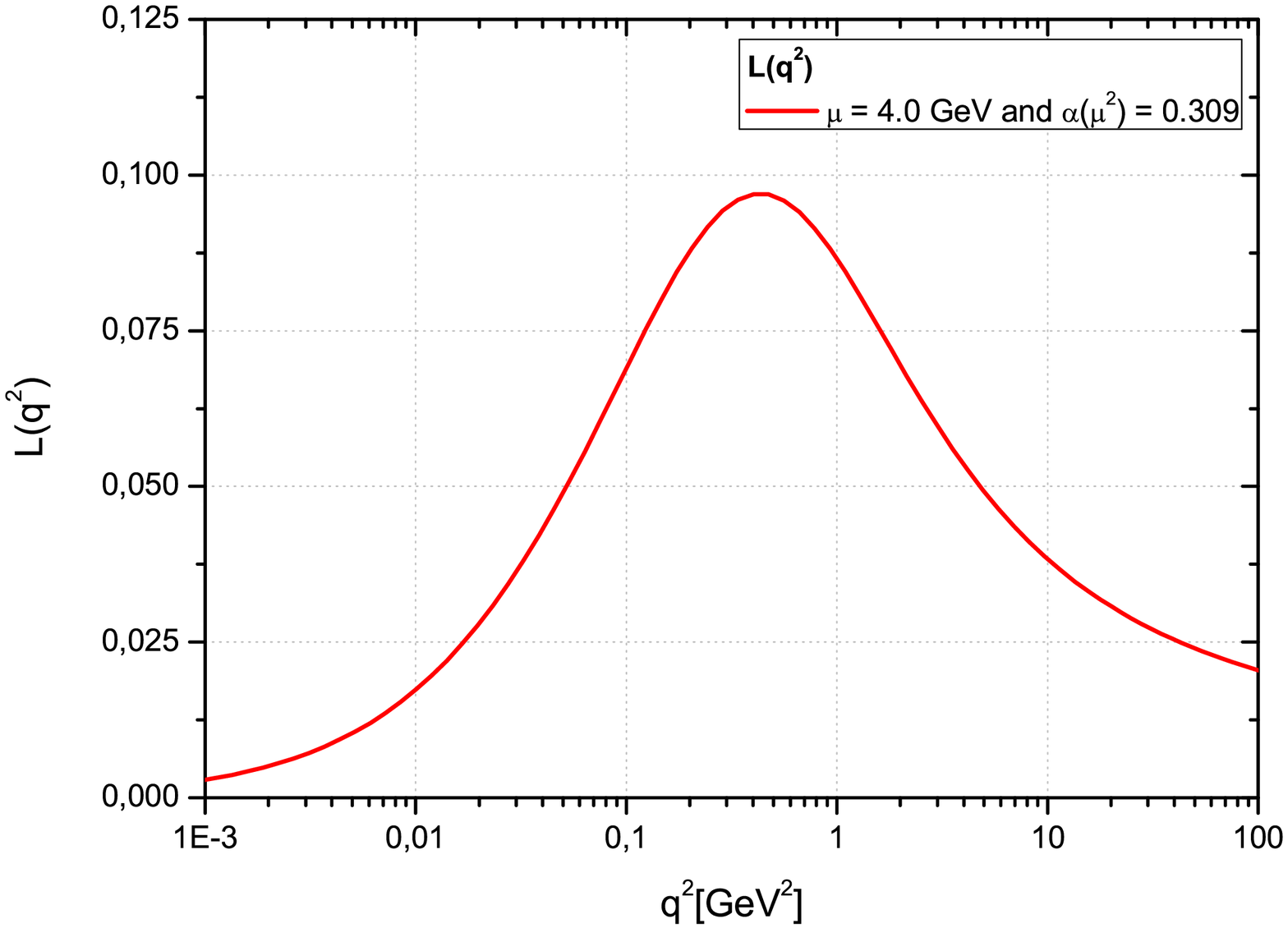} 
\end{minipage}
\vspace{-1.0cm}
\caption{The SDE results for $\Delta(q^2)$, $F(q^2)$, $G(q^2)$, and $L(q^2)$ compared with the corresponding lattice data.}
\label{fig2}
\end{figure}

The solutions obtained for the Green's function of the Eqs.(\ref{SDgl}) and (\ref{simple})
are shown in Fig.~\ref{fig2}, and  are compared with the 
corresponding lattice data \cite{Bogolubsky:2007ud,Sternbeck:2006rd}. Unfortunately, as far we know,
no lattice results exists for $L(q^2)$, and therefore
in the last panel of Fig.~\ref{fig2} we present only our SDE
prediction for it.

Note that, we obtain a good qualitative agreement with the lattice result for $\Delta(q^2)$, $F(q^2)$, and $G(q^2)$. More
specifically, in the case of the gluon propagator, we  clearly  see that both
SDE and lattice results are infrared finite, since \mbox{$\Delta(0)>0$}. Such
feature can be associated to a purely non-perturbative effect
that gives rise to a dynamical gluon mass, which saturates the gluon propagator in the IR. The appearance of 
the gluon mass is also responsible for the infrared finiteness of the  
ghost dressing function, $F(q^2)$, which is clearly shown on the right upper panel
of Fig.~\ref{fig2}  \cite{Aguilar:2008xm,Boucaud:2008ji}. Therefore both, lattice and SDE results, are clearly at odds
with the KO confinement scenario, which requires an enhanced ghost dressing function.
In addition, SDE and lattice~\cite{Sternbeck:2006rd} find no evidence of $G(0)=-1$, which is required  
for the realization of the KO confinement scenario. Specifically, 
the large-volume lattice simulations of~\cite{Sternbeck:2006rd} find that $G(q^2)$ saturates in the deep IR around approximately  $G(0)=-0.6$,  
which is excellent agreement with the value obtained from a recent SDE analysis~\cite{Aguilar:2009pp}. It is important to mention that, very recently, the PT-BFM truncation scheme has been also applied successfully in the case of Yang-Mill in 3d \cite{Aguilar:2010zx}.
%
%%%%%%%%%%%%%%%%%%%%%%%%%%%%%%%%
%    Figure 2
%%%%%%%%%%%%%%%%%%%%%%%%%%%%%%%

Using the all ingredients presented so far, and remembering the fact that the  new PT-BFM Green's functions
satisfy Abelian-like WIs, we may construct the renormalization-group-invariant quantity ${\widehat d}(q^2)$,  defined as \cite{Aguilar:2009nf,Aguilar:2010gm}
\be
{\widehat d}(q^2) = g^2(\mu^2)\frac{\Delta(q^2,\mu^2)}{[1+G(q^2,\mu^2)]^2} \,.
\label{dwg}
\ee 
From ${\widehat d}(q^2)$ we can extract a dimensionless quantity that
corresponds to the non-perturbative generalization of the QCD effective charge, given by
\be
4\pi\alpha(q^2) = [q^2 + m^2(q^2)]{\widehat d}(q^2)\,,
\label{ddef}
\ee
where $m^2(q^2)$ is a momentum-dependent gluon mass, and $\alpha(q^2)=g^2(q^2)/4 \pi$.

\begin{figure}[!t]
\begin{minipage}[b]{0.45\linewidth}
\centering
%\hspace{-1cm}
\includegraphics[scale=0.3]{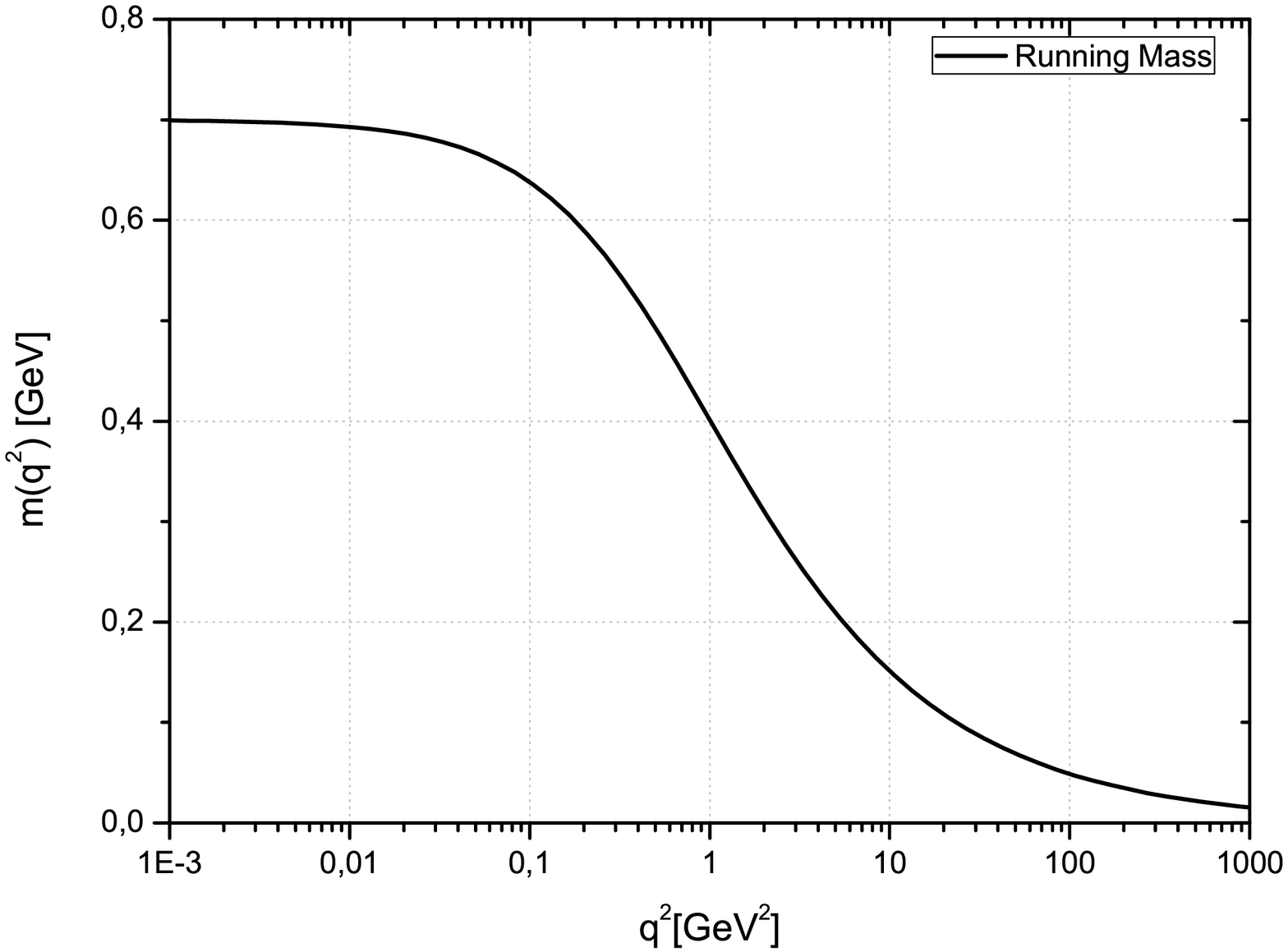}
\end{minipage}
\hspace{0.5cm}
\begin{minipage}[b]{0.50\linewidth}
\centering
%\hspace{-1.5cm}
\includegraphics[scale=0.3]{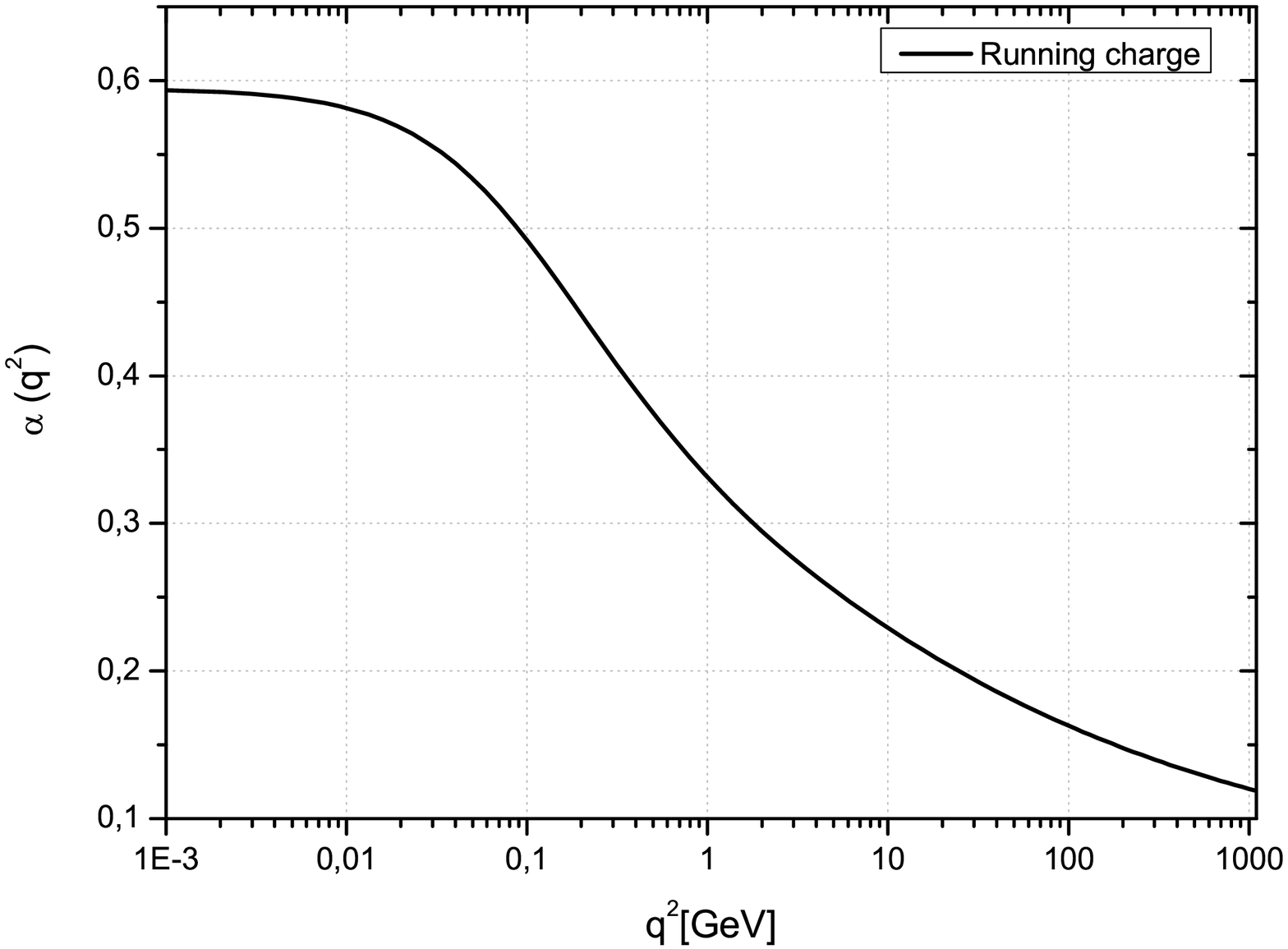}
\end{minipage}
\vspace{-1.0cm}
\caption{ The power-law running mass $m(q^2)$ with $m=700$ MeV (left panel).  The non-perturbative
QCD effective charge, $\alpha(q^2)$,  of Eq.(\ref{ddef}) obtained from the solution of the SDE (right panel).}
\label{fig3}
\end{figure}

Then, assuming a power-law running mass of the type $m^2(q^2)=m^4/(q^2 +m^2)$ \cite{Lavelle:1991ve}(shown in the left panel of Fig.~\ref{fig3}), and 
using the results obtained from the SDE solutions, we obtain the 
effective charge shown in the right panel of Fig.~\ref{fig3}. 
As we can see, $\alpha(q^2)$ saturates at an IR finite value, and displays the correct UV behavior \cite{Aguilar:2009nf,Aguilar:2010gm}.
The presence of the IR fixed point in the behavior of $\alpha(q^2)$ is another manifestation 
of the appearance of the gluon mass \cite{Aguilar:2002tc} that tames the Landau pole, allowing for 
a smooth connection between the IR and UV regions of  $\alpha(q^2)$.
%
%%%%%%%%%%%%%%%%%%%%%%%%%%%%%%%%
%    Figure 3
%%%%%%%%%%%%%%%%%%%%%%%%%%%%%%%

We have presented the basic characteristics of the 
SDEs formulated within of the PT-BFM framework.  We  have seen that the infrared finiteness of the 
gluon propagator and ghost dressing function are associated to the generation of a dynamical gluon mass, which
is also responsible for the appearance appearance of an IR fixed point in the QCD effective charge. In addition, we have shown that
the SDE results for the Green's functions are in nice agreement  with the data of large-volume lattice simulations.

%%%%%%%%%%%%%%%%%%%%%%%%%%%%%%%%%%%%%%%%%%%%%%%%
%% BACKMATTER
%%%%%%%%%%%%%%%%%%%%%%%%%%%%%%%%%%%%%%%%%%%%%%%%
{\bf Acknowledgments:}
The author thanks the organizers of XI Hadrons physics for the pleasure workshop. 
This research is supported by the Brazilian Funding Agency CNPq under the grant 305850/2009-1.

\vspace{-0.5cm}

\bibliographystyle{aipproc}

\begin{thebibliography}{9}


\bibitem{Marciano:1977su}
  W.~J.~Marciano and H.~Pagels,
  %``Quantum Chromodynamics: A Review,''
  Phys.\ Rept.\  {\bf 36}, 137 (1978).
  %%CITATION = PRPLC,36,137;%%


%\cite{Cornwall:1989gv}
\bibitem{Cornwall:1989gv}
  J.~M.~Cornwall and J.~Papavassiliou,
  %``Gauge Invariant Three Gluon Vertex in QCD,''
  Phys.\ Rev.\  D {\bf 40}, 3474 (1989);
  %%CITATION = PHRVA,D40,3474;%%
%\cite{Pilaftsis:1996fh}
%\bibitem{Pilaftsis:1996fh}
  A.~Pilaftsis,
  %``Generalized pinch technique and the background field method in general
  %gauges,''
  Nucl.\ Phys.\  B {\bf 487}, 467 (1997);
%  [arXiv:hep-ph/9607451].
  %%CITATION = NUPHA,B487,467;%%
%\cite{Binosi:2002ft}
%\bibitem{Binosi:2002ft}
  D.~Binosi and J.~Papavassiliou,
  %``The pinch technique to all orders,''
  Phys.\ Rev.\  D {\bf 66}(R), 111901 (2002);
  %[arXiv:hep-ph/0208189].
  %%CITATION = PHRVA,D66,111901;%%
%\cite{Binosi:2009qm}
%\bibitem{Binosi:2009qm}
%  D.~Binosi and J.~Papavassiliou,
  %``Pinch Technique: Theory and Applications,''
  Phys.\ Rept.\  {\bf 479}, 1 (2009).
%  [arXiv:0909.2536 [hep-ph]].
  %%CITATION = PRPLC,479,1;%%
  
%\cite{Abbott:1980hw}
\bibitem{Abbott:1980hw}
L.~F.~Abbott,
%``The Background Field Method Beyond One Loop,''
Nucl.\ Phys.\ B {\bf 185}, 189 (1981).
%%CITATION = NUPHA,B185,189;%%
  

%\cite{Cucchieri:2007md}
\bibitem{Cucchieri:2007md}
  A.~Cucchieri and T.~Mendes,
  %``What's up with IR gluon and ghost propagators in Landau gauge? A puzzling
  %answer from huge lattices,''
  PoS {\bf LAT2007}, 297 (2007);
%  [arXiv:0710.0412 [hep-lat]].
  %%CITATION = POSCI,LAT2007,297;%%
%\cite{Cucchieri:2007rg}
%\bibitem{Cucchieri:2007rg}
 % A.~Cucchieri and T.~Mendes,
%``Constraints on the IR behavior of the gluon propagator in Yang-Mills
  %theories,''
  Phys.\ Rev.\ Lett.\  {\bf 100}, 241601 (2008);
%  [arXiv:0712.3517 [hep-lat]].
  %%CITATION = PRLTA,100,241601;%%
%\cite{Cucchieri:2009zt}
%\bibitem{Cucchieri:2009zt}
  A.~Cucchieri and T.~Mendes,
  %``Landau-gauge propagators in Yang-Mills theories at beta = 0: massive
  %solution versus conformal scaling,''
  Phys.\ Rev.\  D {\bf 81}, 016005 (2010);
%  [arXiv:0904.4033 [hep-lat]].
  %%CITATION = PHRVA,D81,016005;%%
%\cite{Oliveira:2009nn}
%\bibitem{Oliveira:2009nn}
  O.~Oliveira and P.~J.~Silva,
  %``The lattice infrared Landau gauge gluon propagator: from finite volume to
  %the infinite volume,''
  arXiv:0911.1643 [hep-lat].
  %%CITATION = ARXIV:0911.1643;%%


%\cite{Bogolubsky:2007ud}                        
\bibitem{Bogolubsky:2007ud}
  I.~L.~Bogolubsky {\it et al.},
  %``The Landau gauge gluon and ghost propagators in 4D SU(3) gluodynamics in
  %large lattice volumes,''
  PoS {\bf LAT2007}, 290 (2007);
%  [arXiv:0710.1968 [hep-lat]].
  %%CITATION = POSCI,LAT2007,290;%%
%\cite{Bogolubsky:2009dc}
%\bibitem{Bogolubsky:2009dc}
  %I.~L.~Bogolubsky, E.~M.~Ilgenfritz, M.~Muller-Preussker and A.~Sternbeck,
  %``Lattice gluodynamics computation of Landau gauge Green's functions in the
  %deep infrared,''
  Phys.\ Lett.\  B {\bf 676}, 69 (2009).
  %[arXiv:0901.0736 [hep-lat]].
  %%CITATION = PHLTA,B676,69;%%
 
 
 %\cite{Sternbeck:2006rd}              
\bibitem{Sternbeck:2006rd}
  A.~Sternbeck,
  %``The infrared behavior of lattice QCD Green's functions,''
  arXiv:hep-lat/0609016.
  %%CITATION = HEP-LAT/0609016;%%

%\cite{Cornwall:1982zr}                                       
\bibitem{Cornwall:1982zr}
J.~M.~Cornwall,
%``Dynamical Mass Generation In Continuum QCD,''
Phys.\ Rev.\ D {\bf 26}, 1453 (1982). 
%%CITATION = PHRVA,D26,1453;%%

%\cite{Aguilar:2006gr}
\bibitem{Aguilar:2006gr}
  A.~C.~Aguilar and J.~Papavassiliou,
  %``Gluon mass generation in the PT-BFM scheme,''
  JHEP {\bf 0612}, 012 (2006);
%  [arXiv:hep-ph/0610040].
  %%CITATION = JHEPA,0612,012;%%
%\cite{Binosi:2008qk}
%\bibitem{Binosi:2008qk}
  D.~Binosi and J.~Papavassiliou,
  %``New Schwinger-Dyson equations for non-Abelian gauge theories,''
  JHEP {\bf 0811}, 063 (2008); 
%  [arXiv:0805.3994 [hep-ph]].
  %%CITATION = JHEPA,0811,063;%%
%\cite{Binosi:2007pi}
%\bibitem{Binosi:2007pi}
%  D.~Binosi and J.~Papavassiliou,
  %``Gauge-invariant truncation scheme for the Schwinger-Dyson equations of
  %QCD,''
  Phys.\ Rev.\  D {\bf 77}, 061702 (2008)
 % [arXiv:0712.2707 [hep-ph]].
  %%CITATION = PHRVA,D77,061702;%%

%\cite{Aguilar:2008xm}
\bibitem{Aguilar:2008xm}
  A.~C.~Aguilar, D.~Binosi and J.~Papavassiliou,
  %``Gluon and ghost propagators in the Landau gauge: Deriving lattice results
  %from Schwinger-Dyson equations,''
  Phys.\ Rev.\  D {\bf 78}, 025010 (2008).
%  [arXiv:0802.1870 [hep-ph]].
  %%CITATION = PHRVA,D78,025010;%%    





\bibitem{Grassi:1999tp}
  P.~A.~Grassi, T.~Hurth and M.~Steinhauser,
  %``Practical algebraic renormalization,''
  Annals Phys.\  {\bf 288}, 197 (2001);
%  [arXiv:hep-ph/9907426].
  %%CITATION = APNYA,288,197;%%
%\cite{Binosi:2002ez}
%\bibitem{Binosi:2002ez}
  D.~Binosi and J.~Papavassiliou,
  %``Pinch technique and the Batalin-Vilkovisky formalism,''
  Phys.\ Rev.\ D {\bf 66}, 025024 (2002).
%  [arXiv:hep-ph/0204128].
  %%CITATION = HEP-PH 0204128;%%
  
%\cite{Kugo:1979gm}
\bibitem{Kugo:1979gm}
  T.~Kugo and I.~Ojima,
  %``Local Covariant Operator Formalism Of Nonabelian Gauge Theories And Quark
  %Confinement Problem,''
  Prog.\ Theor.\ Phys.\ Suppl.\  {\bf 66}, 1 (1979).
  %%CITATION = PTPSA,66,1;%%
 
 %\cite{Aguilar:2009pp}
\bibitem{Aguilar:2009pp}
  A.~C.~Aguilar, D.~Binosi and J.~Papavassiliou,
  %``Indirect determination of the Kugo-Ojima function from lattice data,''
  JHEP {\bf 0911}, 066 (2009).
%  [arXiv:0907.0153 [hep-ph]].
  %%CITATION = JHEPA,0911,066;%%
  

%\cite{Aguilar:2009nf}
\bibitem{Aguilar:2009nf}
 A.~C.~Aguilar {\it et al.},
  %``Non-perturbative comparison of QCD effective charges,''
  Phys.\ Rev.\  D {\bf 80}, 085018 (2009).
%  arXiv:0906.2633 [hep-ph].
  %%CITATION = ARXIV:0906.2633;%%

%\cite{Grassi:2004yq}
\bibitem{Grassi:2004yq}
  P.~A.~Grassi, T.~Hurth and A.~Quadri,
  %``On the Landau background gauge fixing and the IR properties of YM Green
  %functions,''
  Phys.\ Rev.\  D {\bf 70}, 105014 (2004).
%  [arXiv:hep-th/0405104].
  %%CITATION = PHRVA,D70,105014;%% 

%\cite{Boucaud:2008ji}
\bibitem{Boucaud:2008ji}            
Ph.~Boucaud {\it et al.},
%``IR finiteness of the ghost dressing function from numerical resolution of
%the ghost SD equation,''
JHEP {\bf 0806} (2008) 012.
%[arXiv:0801.2721 [hep-ph]];
%%CITATION = JHEPA,0806,012;%%
  
 
%\cite{Aguilar:2010zx}
\bibitem{Aguilar:2010zx}
  A.~C.~Aguilar, D.~Binosi and J.~Papavassiliou,
  %``Nonperturbative gluon and ghost propagators for d=3 Yang-Mills,''
  arXiv:1004.2011 [hep-ph].
  %%CITATION = ARXIV:1004.2011;%%

 
%\cite{Aguilar:2010gm}                       
\bibitem{Aguilar:2010gm}
  A.~C.~Aguilar, D.~Binosi and J.~Papavassiliou,
  %``QCD effective charges from lattice data,''
  arXiv:1004.1105 [hep-ph].
  %%CITATION = ARXIV:1004.1105;%%
 
 



%\cite{Lavelle:1991ve}
\bibitem{Lavelle:1991ve}
  M.~Lavelle,
  %``Gauge invariant effective gluon mass from the operator product expansion,''
  Phys.\ Rev.\  D {\bf 44}, 26 (1991);
  %%CITATION = PHRVA,D44,26;%%
%\cite{Dudal:2008sp}
%\bibitem{Dudal:2008sp}
D.~Dudal {\it et al.},
%``A refinement of the Gribov-Zwanziger approach in the Landau gauge: infrared
%propagators in harmony with the lattice results,''
Phys.\ Rev.\  D {\bf 78}, 065047 (2008).  
%[arXiv:0806.4348 [hep-th]].
%%CITATION = PHRVA,D78,065047;%%

%\cite{Aguilar:2002tc}
\bibitem{Aguilar:2002tc}
  A.~C.~Aguilar, A.~A.~Natale and P.~S.~Rodrigues da Silva,
  %``Relating a gluon mass scale to an infrared fixed point in pure gauge QCD,''
  Phys.\ Rev.\ Lett.\  {\bf 90}, 152001 (2003).
%  [arXiv:hep-ph/0212105].
  %%CITATION = PRLTA,90,152001;%%



 



\end{thebibliography}

\end{document}